\documentclass{jfm}
\usepackage{graphicx}
\usepackage{epstopdf, epsfig}
\usepackage{float}
\usepackage{amsmath}
\usepackage{graphicx}
\usepackage{caption}
\usepackage{subcaption}
\usepackage{natbib}

\shortauthor{S. A. Reijers, J. H. Snoeijer and H. Gelderblom}
\title{Droplet deformation by short laser-induced pressure pulses}

\author{Sten A. Reijers\aff{1}
  \corresp{\email{s.a.reijers@utwente.nl}},
 Jacco H. Snoeijer\aff{1,2}
  \and Hanneke Gelderblom\aff{1}}

\affiliation{\aff{1}Physics of Fluids Group, Faculty of Science and Technology, MESA+ Institute, University of Twente, P.O. Box 217, 7500 AE Enschede, The Netherlands
\aff{2}Mesoscopic Transport Phenomena, Eindhoven University of Technology, Den Dolech 2, 5612 AZ Eindhoven, The Netherlands}

\begin{document}

\maketitle

\begin{abstract}
When a free-falling liquid droplet is hit by a laser it experiences a strong ablation driven pressure pulse. Here we study the resulting droplet deformation in the regime where the ablation pressure duration is short, i.e. comparable to the time scale on which pressure waves travel through the droplet. To this end an acoustic analytic model for the pressure-, pressure impulse- and velocity fields inside the droplet is developed in the limit of small density fluctuations. This model is used to examine how the droplet deformation depends on the pressure pulse duration while the total momentum to the droplet is kept constant. Within the limits of this analytic model, we demonstrate that when the total momentum transferred to the droplet is small the droplet shape-evolution is indistinguishable from an incompressible droplet deformation. However, when the momentum transfer is increased the droplet response is strongly affected by the pulse duration. In this later regime, compressed flow regimes alter the droplet shape evolution considerably.
\end{abstract}
\section{Introduction}
\begin{figure}
  \centerline{\includegraphics[width=0.9\linewidth]{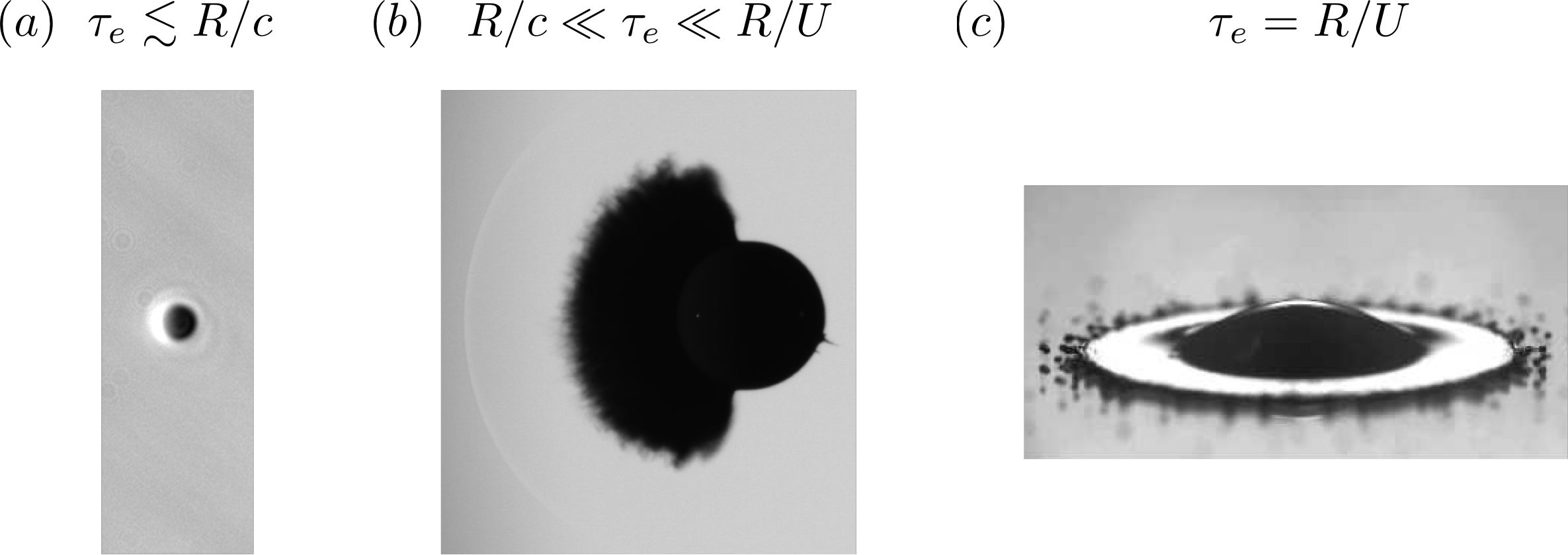}}
  \caption{An illustration of three different impact duration regimes on a droplet. (a) A nanosecond laser pulse impacting from the left on a micron-sized liquid tin droplet leads to plasma formation (white glow in the image) and subsequently plasma-mediated ablation of the droplet \citep{Kurilovich2016}. The typical ablation pressure duration is comparable to the plasma-decay duration, which is of the order of the acoustic time scale:  $\tau_e \lesssim R/ c$, where $R$ is the initial droplet radius and $c$ the speed of sound inside the droplet (image taken from \citet{Kurilovich2016}). (b) Impact of a nanosecond laser pulse onto a millimetre-sized dyed water droplet can lead to vaporization and mist cloud formation, the accompanying shock wave in the surrounding air is also visible \citep{Klein2015}. The typical vapour-recoil induced ablation pressure duration is much longer than the acoustic time scale, but much shorter than the time scale on which the droplet deforms: $R/c \ll \tau_e \ll R/U$, where $U$ is the propulsion speed of the droplet  (image taken $5 \mu s$ after laser impact, image courtesy A.L. Klein). (c) For the impact of a droplet onto a solid surface the typical interaction time is equal to the deformation time  $\tau_e = R/U$, where $U$ is the impact speed of the droplet (image taken from \citet{Josserand2016}).}
\label{fig:initialimage}
\end{figure}

The impact of a short laser pulse onto a free-falling absorbing liquid droplet induces a rapid phase change in a thin superficial layer on the illuminated side of the droplet \citep{Klein2015, Kurilovich2016}. The resulting vaporization, explosive boiling or even plasma formation gives rise to mass ablation; see figures \ref{fig:initialimage}a and b. Subsequently, a recoil pressure wave propagates into the droplet and causes a net momentum transfer \citep{Sigrist1978,Apitz2005,Klein2015}. As a consequence the droplet is propelled forward and strongly deforms  \citep{Klein2015, Gelderblom2016}. However, the way in which these pressure waves establish inside the droplet over time, which is in particular relevant for short pulse durations, has so far remained unexplored.

In this study we aim to understand the fluid dynamic response of a droplet to a short ablation driven pressure pulse. Next to this ablation pressure, a laser impact could trigger pressure waves inside the droplet through a number of other mechanisms \citep{Sigrist1986}. Electrostriction and radiation pressures are of negligible influence compared to the ablation pressure \citep{Sigrist1986}. However, the local heating of the liquid close to the droplet surface can induce significant thermoelastic waves that result from thermal expansion \citep{Sigrist1978,Wang2001}. Furthermore, for high laser intensities dielectric breakdown on the droplet surface can lead to the generation of shock waves inside the droplet \citep{Zhang1987,Vogel1996,Lauterborn2013} or even plasma generation inside a transparent droplet \citep{Lindinger2004, Geints2010, Avila2016}. These mechanisms could have a strong influence on the droplet response. Indeed, cavitation phenomena, shock waves and rapid interface acceleration can give rise to fast jetting, bubble collapse and interfacial instabilities \citep{Vogel1996, Thoroddsen2009, Sun2009, Tagawa2012,Avila2016}. The study of these violent, highly non-linear response regimes is beyond the scope of the present study. Instead, we examine how an ablation pressure pulse is communicated throughout the droplet and triggers droplet deformation.

An important application of laser-induced droplet deformation is found in Laser Produced Plasma light-sources to generate Extreme Ultra Violet (EUV) light used for nanolithography \citep{Fujioka2008, Banine2011}. In these sources small tin droplets are converted into a plasma by a two-stage laser impact process \citep{Banine2011}. Upon the first impact, the droplet deforms into a thin flat sheet which is thereafter ionized by a second more powerful laser. A key question to improve this source is how the droplet deformation changes when the laser pulse duration is shortened. 

Up to now, the response of a droplet due to a laser impact has been studied by using incompressible hydrodynamics to model the droplet deformation \citep{Klein2015,Gelderblom2016}. In these models the interaction of the laser with the droplet is described by an ablation pressure $p_e$ acting on the surface of the droplet for a duration $\tau_e$. The impulse $p_e\tau_e$ resulting from this ablation pressure causes a momentum transfer to the droplet $\rho_0 R^3 U$, where $\rho_0$ is the liquid density, $R$ the initial droplet radius and $U$ the center-of-mass speed, which therefore scales as \citep{Gelderblom2016}
\begin{align}
U \sim \frac{p_e \tau_e}{\rho_0 R}.
\label{eq:u}
\end{align}
The deformation in these incompressible models is calculated by a pressure impulse approach that is also used for studies on the impact of liquid bodies onto solids \citep{Batchelor1967, Cooker1995, Antkowiak2007}. As long as the duration of the ablation pressure is long compared to the acoustic time scale $R/c$, where $c$ is the speed of sound inside the droplet, and the amplitude $p_e$ is such that no shockwaves are created, these incompressible models are valid and the droplet response can be considered incompressible \citep{Gelderblom2016}. For example, for classical droplet impact onto a solid the deformation time scale $\tau_i = R/U$ is of the same order as the impact duration $\tau_e$ which is much longer than $R/c$ (see e.g. \citet{Clanet2004, Josserand2016}), as illustrated in figure \ref{fig:initialimage}c. 

By contrast, the impact of a laser pulse provides a means to shorten the duration of the ablation pressure considerably, and thereby to transfer the same amount of momentum $p_e\tau_e$ to the droplet in a shorter time. The ablation-pressure duration can for example be decreased by increasing the laser pulse energy to move to the plasma-mediated ablation regime, which leads to more violent and shorter lived ablation pressures \citep{Kurilovich2016}, as illustrated in figure \ref{fig:initialimage}a. A further decrease of the ablation-pressure duration can be obtained by directly shortening the laser-pulse duration \citep{Chichkov1996}. In these cases $\tau_e$ is shortened significantly such that it becomes comparable to or even smaller than $R/c$ such that the droplet response is compressible and incompressible models breakdown. We note that for laser-induced ablation $\tau_e \ll \tau_i$ such that the droplet remains undeformed during impact \citep{Gelderblom2016}. Indeed, in figure \ref{fig:initialimage}b we observe that the mist cloud resulting from mass ablation acts on the surface of an undeformed droplet.  

In this paper we study the response of a droplet to a short ablation pressure acting on its surface. In particular, we focus on the question how the droplet deformation dynamics depends on the ablation pressure duration at fixed impulse in the regime where $\tau_e \lesssim R/c$. Hence we consider the situation where the pressure field inside the droplet is not yet established during the pressure pulse and the droplet response is no longer incompressible. To this end, we develop a linearly compressible analytic model for the droplet response to short pressure pulses. In \S \ref{section:problemformulationandmethods} we introduce the analytic model and discuss the regime in which it applies. In \S \ref{section:resultssection} we first compare our analytic results to a compressible lattice-Boltzmann simulation. Next we use the analytic model to study the effects of shortening the pulse duration at constant impulse on the pressure-, pressure impulse-, velocity- and deformation fields of the droplet.

\section{Problem formulation \& methods}
\label{section:problemformulationandmethods}
In this section we derive a model to describe the spatio-temporal response of a droplet to an ablation pressure acting on its surface. In \S \ref{section:scalinganalysis} we provide a scaling analysis to delineate three different regimes in the response of the droplet to this pressure pulse. Analytic expressions for the pressure and velocity fields inside the droplet as function of the pressure pulse are derived in \S \ref{section:weaklycompressiblemodel}. Finally in \S \ref{section:lbm} we discuss the lattice-Boltzmann method that we use to support our analytic findings.

\subsection{Scaling analysis}\label{section:scalinganalysis}
\begin{figure}
  \centerline{\includegraphics[width=0.5\linewidth]{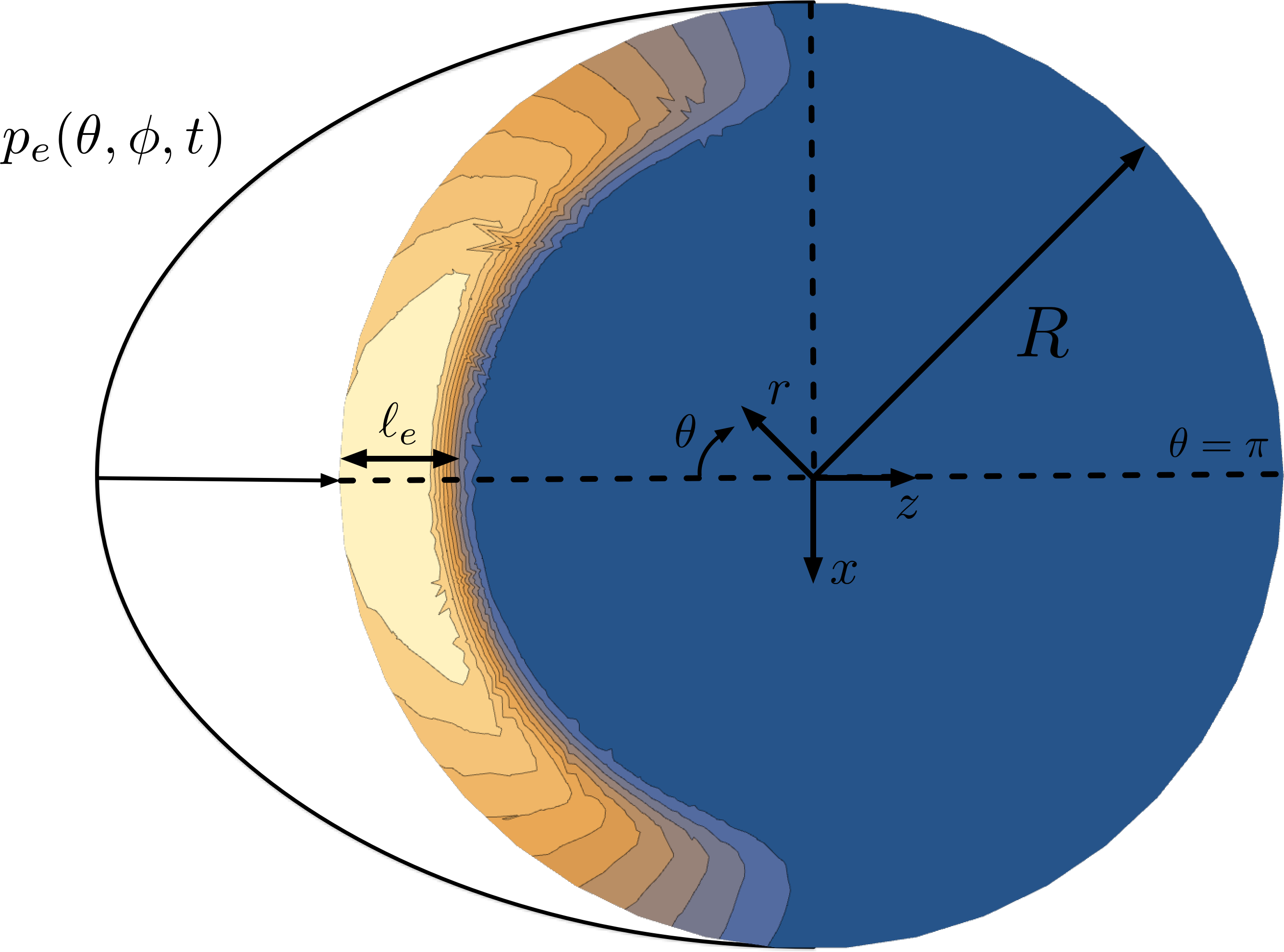}}
  \caption{Sketch of the problem: an ablation pressure of amplitude $p_e(\theta,\phi,t)$ and duration $\tau_e$ acts on the surface of a droplet with radius $R$. As a result, a pressure field is induced inside the droplet over a depth $\ell_e\sim c\tau_e$.  The colorbar denotes the pressure amplitude (blue is ambient and yellow is peak pressure). The spherical coordinate system ($r$,$\theta$,$\phi$) used is indicated (the azimuthal angle $\phi$ is not shown but rotates around the z-axis).}
\label{fig:geometry}
\end{figure}
\label{section:physicalpicture}
We consider a spherical droplet with radius $R$ and density $\rho_0$ that is submitted to an ablation pressure on the illuminated side with an amplitude $p_e$ and a duration $\tau_e$, see figure \ref{fig:geometry}. The total impulse received by the droplet is given by $J \sim p_e \tau_e$. In this work we explore the effect of decreasing the pulse duration while keeping the total impulse transferred to the droplet constant; i.e. decreasing $\tau_e$ at constant $J$.

During the pulse the pressure disturbance on the surface of the droplet penetrates over a length-scale $\ell_e\sim c \tau_e$, where $c$ is the speed of sound in the droplet. If $\ell_e$ is short compared to $R$ all momentum is initially concentrated inside a thin layer, as is illustrated in figure \ref{fig:geometry}. By contrast, if $\ell_e > R$, all fluid inside the droplet has experienced a change in momentum directly after the pulse. The ratio between $\ell_e$ and $R$ is quantified by the acoustic Strouhal number and is a dimensionless pressure pulse duration
\begin{equation}
\text{St} = \frac{\ell_e}{R} = \frac{c \tau_e}{R}. 
\end{equation}
To investigate the effect of short pulse durations, we are interested in the limit  $\text{St} \lesssim 1$. 

When $\tau_e$ is decreased at constant $J$, $p_e$ rises. From momentum conservation  in this thin layer it follows that the typical velocity induced inside $\ell_e$ is given by $u_e\sim p_e/(\rho_0 c)$, where $\rho_0$ is the density of the liquid droplet. Hence we observe that a large $p_e$ induces large velocities in $\ell_e$, which is quantified by the acoustic Mach number and is a dimensionless pressure pulse amplitude
\begin{equation}
\text{Ma} = \frac{u_e}{c} = \frac{p_e}{p_0}, 
\end{equation}
where $p_0 = \rho_0 c^2$. When $\text{Ma}$ is large the fluid response inside the droplet is non-linear and shock waves dominate the flow. If $\text{Ma}$ small, the flow inside the droplet can be considered linear.

The product $\text{Ma}\text{St}$ sets the total dimensionless impulse received by the droplet
\begin{equation}
\text{St}\text{Ma} = \frac{p_e\tau_e}{\rho_0 R c} = \frac{U}{c},
\end{equation}
where (\ref{eq:u}) is used to express the center-of-mass velocity $U$ of the droplet as whole. This product is often referred to as the global Mach number of the droplet.

\begin{figure}
  \centerline{\includegraphics[width=0.75\linewidth]{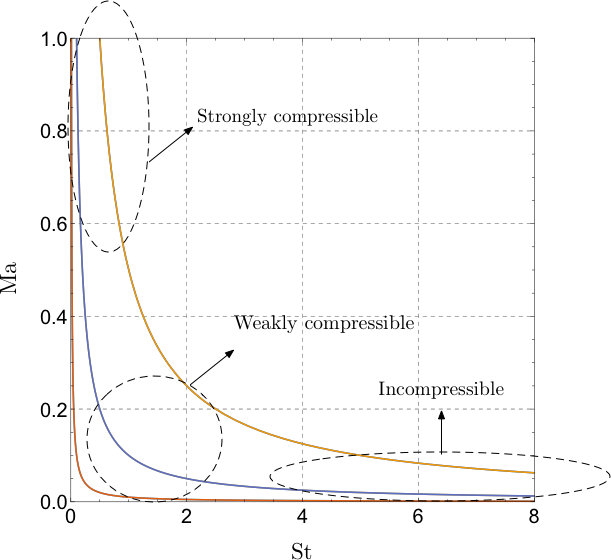}}
  \caption{A phase diagram showing lines of constant impulse transfer to the droplet. The red line is the isoline $\text{St}\text{Ma}=0.01$, the blue line is the isoline $\text{St}\text{Ma}=0.1$ and the yellow line is the isoline $\text{St}\text{Ma}=0.5$. The plot shows three distinct regimes can be observed at constant impulse: a strongly compressible regime, a weakly compressible regime and an incompressible regime. The weakly compressible regime is the focus of the present work.}
\label{fig:differentregimeswithconstantimpulse}
\end{figure}

One can use $\text{Ma}$ and $\text{St}$ to delineate different regimes in the droplet response, as illustrated in figure \ref{fig:differentregimeswithconstantimpulse}. For lines of constant $\text{Ma}\text{St}$ (hence constant impulse), we can identify three regimes. Firstly, when $\text{St}$ is small and $\text{Ma}$ is large, we are in a strongly compressible regime where nonlinear advective acceleration and nonlinear viscous dampening need to be taken into account to describe the flow. Secondly, for intermediate $\text{Ma}$ and $\text{St}$, compressible effects are important but nonlinear effects are small, which renders this regime analytically accessible. This regime, which we term the weakly compressible regime, will be the main focus of this paper. Finally, when $\text{St}\gg 1$ and $\text{Ma} \ll 1$, we enter the incompressible regime that was subject of previous studies where long pulse durations (large $\text{St}$) were considered \citep{Gelderblom2016}. This regime is also relevant to droplet impact studies on rigid surfaces, see e.g.~\citet{Yarin2006,Clanet2004,Richard2002, Philippi2016,Wildeman2016,Josserand2016}.
\subsection{The weakly compressible model}\label{section:weaklycompressiblemodel}
In the weakly compressible regime, i.e.~for intermediate Ma, St (see figure \ref{fig:differentregimeswithconstantimpulse}), we can expand the pressure $p$, density $\rho$ and velocity $\boldsymbol{u}$ as a constant plus a small oscillatory part, analogous to \citet[p166]{Batchelor1967} 
\begin{align}
p(\boldsymbol{x},t) = p_0 + p_1(\boldsymbol{x},t),\nonumber\\
\rho(\boldsymbol{x},t) = \rho_0 + \rho_1(\boldsymbol{x},t),\label{eq:hydrodynamicquantities}\\
\boldsymbol{u}(\boldsymbol{x},t) = \boldsymbol{u}_1(\boldsymbol{x},t),\nonumber
\end{align}
where $t$ is the time, $\boldsymbol{x}=(r,\theta,\phi)$ are the spherical coordinates defined in figure \ref{fig:geometry} and $\boldsymbol{u}_0 = \boldsymbol{0}$. The pressure can formally be expressed as $p = p(\rho, S, T)$, where $S$ is the entropy and $T$ is the temperature. However, in order to simplify the analysis we only take into account the density fluctuations on first order
\begin{align}
p(\boldsymbol{x},t) = p_0 + c^2\rho_1(\boldsymbol{x},t).
\label{eq:equationofstateideal}
\end{align}
Since we are interested in the flow inside the droplet directly after the pulse we introduce the following non-dimensionalisation, following the scaling analysis of the previous paragraph
\begin{equation}
  \boldsymbol{u}= \frac{p_e}{\rho_0 c}\boldsymbol{\tilde{u}},\quad  \boldsymbol{x}=R\boldsymbol{\tilde{x}},\quad t = \tau_e\tilde{t},\quad p = p_e\tilde{p},\quad \rho=\frac{p_e}{c^2}\tilde{\rho}.
\label{eq:scaling}
\end{equation}
where the tildes refer to the dimensionless parameters. From now on we drop the tildes and work with the dimensionless parameters. The linearized continuity equation and linearized momentum equation are given by
\begin{align}
\frac{\partial p_1}{\partial t} + \text{St}\left(\boldsymbol{\nabla}\cdot\boldsymbol{u}_1\right) &= 0,\label{eq:massconservationperturbed}\\
\frac{1}{\text{St}}\frac{\partial \boldsymbol{u}_1}{\partial t} &= -\boldsymbol{\nabla}p_1 + \frac{1}{\text{Re}}\nabla^2\boldsymbol{u}_1+\frac{1}{\text{Re}_v}\boldsymbol{\nabla}(\boldsymbol{\nabla}\cdot\boldsymbol{u}_1),\label{eq:momentumconservationperturbed}
\end{align}
where $\text{Re} = \rho_0 R c /\mu$ is the Reynolds number with $\mu$ the dynamic viscosity and $\text{Re}_v = \rho_0 R c/(\frac{1}{3}\mu+\kappa)$ the Reynolds number for volume changes, where $\kappa$ is the bulk viscosity. Although the Reynolds number in experiments is typically large ($\text{Re}\sim 10^3$) we will see later on that we need to retain the viscous terms in (\ref{eq:momentumconservationperturbed}) to overcome singularities when converging pressure waves superimpose in the center of the droplet. We note that the equations do not depend on $\text{Ma}$, since essentially this is a low order Mach expansion of the compressible Navier-Stokes equations. By taking the divergence of (\ref{eq:momentumconservationperturbed}) and using (\ref{eq:massconservationperturbed}) we obtain a viscous wave equation for the acoustic field inside the droplet
\begin{align}
\frac{\partial^2 p_1}{\partial t^2} -\text{St}^2\nabla^2 p_1 = \frac{1}{\text{Re}_a}\left[\nabla^2\frac{\partial p_1}{\partial t}\right],\label{eq:waveequviscousperturbed}
\end{align}
where $\frac{1}{\text{Re}_a} =\text{St}\left(\frac{1}{\text{Re}}+\frac{1}{\text{Re}_v}\right)$ is an effective Reynolds number for the viscous dissipation in the acoustic wave \citep[p97]{Blackstock2000}. Below we describe how this equation for $p_1(\boldsymbol{x},t)$ is solved for the problem at hand. In \S \ref{section:velocityfielddefinition} we show how the velocity field $\boldsymbol{u}_1(\boldsymbol{x},t)$ can be computed once $p_1(\boldsymbol{x},t)$ is known.

\subsubsection{The pressure field}
We solve (\ref{eq:waveequviscousperturbed}) subject to a pressure boundary condition on the droplet surface. At the interface of the droplet the pressure must be continuous, since surface tension effects are negligible on the acoustic time scale. We assume that the magnitude of the pressure variations in the gas phase are much smaller than those inside the droplet, since the density and the viscosity of the gas phase are much smaller. Therefore, the stress in the gas phase may be considered constant and equal to $p_0$. As a consequence, the oscillatory part of the pressure at the surface must satisfy
\begin{align}
p_1(1,t) =0,
\label{eq:boundaryconditionfreesurface}
\end{align}
where we have assumed that the interface remains immobile and therefore the droplet spherical during the pulse. This assumption is justified when the pulse duration $\tau_e$ is much smaller than the typical interface deformation time scale $\tau_{int} = R/u_e$, or in dimensionless form $\text{St}\text{Ma} \ll 1$. Typically in experiments $\text{St}\text{Ma} \sim 10^{-2}-10^{-1}$. An ablation pressure acting on the surface of the droplet will be introduced through a Green's function formalism \citep[Chapter 7]{MorseFeshbach1953}.
\begin{figure}
  \centerline{\includegraphics[width=0.9\linewidth]{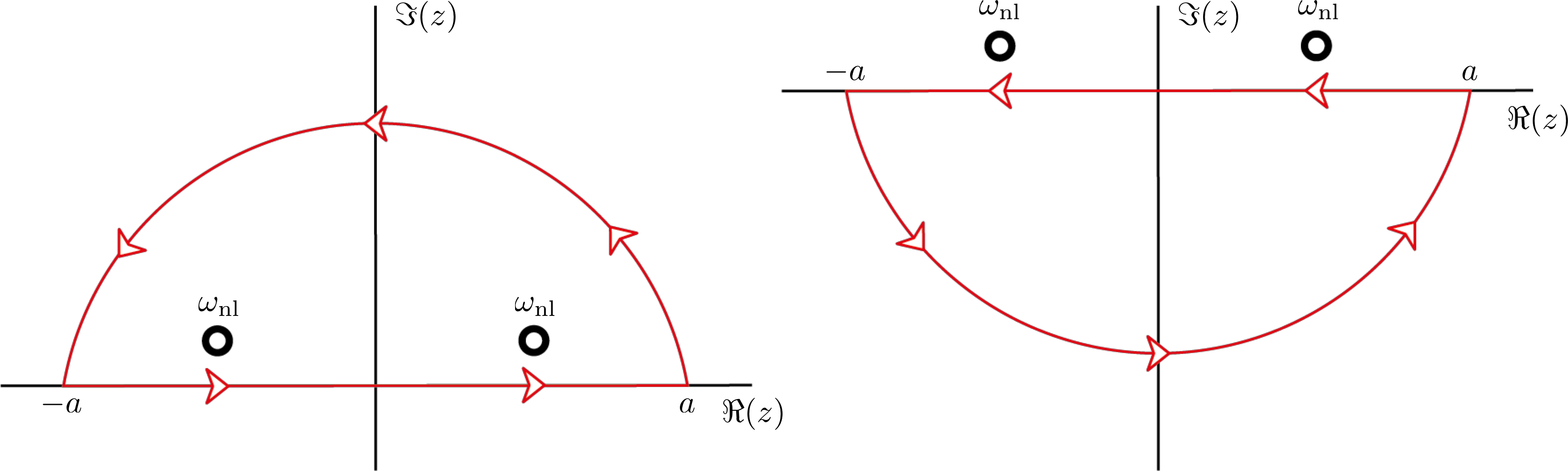}}
  \caption{Jordan curves used to evaluate the inverse Fourier transform (\ref{eq:greensfunctionviscidtransformed}) in the complex plane where $a \to\infty$. (a) contour used for $t>t_0$, which includes the poles. (b) contour used for $t<t_0$ to obey the causality condition for the Green's function.}
\label{fig:complexintegralviscid}
\end{figure}
The general solution for the spatio-temporal pressure field inside the droplet is given by
\begin{align}
p_1(\boldsymbol{x},t) &=  \iiint\limits_{V} \bigg[\frac{\partial G(\boldsymbol{x},t;\boldsymbol{x}_0,t_0)}{\partial t_0} p_1(\boldsymbol{x_0},t_0) - \bigg(\frac{\partial p_1(\boldsymbol{x}_0,t_0)}{\partial t_0} -\frac{p_1(\boldsymbol{x}_0,t_0)}{\text{Re}_a}\nabla^2_{\boldsymbol{x}_0} \bigg)G(\boldsymbol{x},t;\boldsymbol{x}_0,t_0)\bigg]\bigg|_{t_0=0}^{t^+} \nonumber \\& \,\mathrm{d}V_0 +\int_0^{t^+} dt_0 \oint\limits_{S} \bigg[ G(\boldsymbol{x},t;\boldsymbol{x}_0,t_0)\bigg(\text{St}^2\boldsymbol{\nabla}_{\boldsymbol{x}_0}p_1(\boldsymbol{x}_0,t_0) \nonumber+\frac{1}{\text{Re}_a}\boldsymbol{\nabla}_{\boldsymbol{x}_0}\frac{\partial p_1(\boldsymbol{x}_0,t_0)}{\partial t_0}\bigg) \\&- \boldsymbol{\nabla}_{\boldsymbol{x}_0}G(\boldsymbol{x},t;\boldsymbol{x}_0,t_0)\bigg(\text{St}^2 p_1(\boldsymbol{x}_0,t_0) + \frac{1}{\text{Re}_a}\frac{\partial p_1(\boldsymbol{x}_0,t_0)}{\partial t_0}\bigg)\bigg]\cdot \boldsymbol{n}\,\mathrm{d}S_0 \label{eq:greensfunctiongeneralsolutionviscid},
\end{align}
where $G(\boldsymbol{x},t;\boldsymbol{x}_0,t_0)$ is the Green's function satisfying
\begin{align}
\frac{\partial^2 G(\boldsymbol{x},t;\boldsymbol{x}_0,t_0)}{\partial t^2} -\text{St}^2\nabla^2 G(\boldsymbol{x},t;\boldsymbol{x}_0,t_0) &- \frac{1}{\text{Re}_a}\left[\nabla^2\frac{\partial G(\boldsymbol{x},t;\boldsymbol{x}_0,t_0)}{\partial t}\right] =\delta(\boldsymbol{x}-\boldsymbol{x}_0)\delta(t- t_0),\label{eq:waveequviscousperturbedgreensfunc}
\end{align}
where we use a spherical coordinate system for $\boldsymbol{x}$ and $\boldsymbol{x}_0$. To find the general solution to (\ref{eq:waveequviscousperturbedgreensfunc}), we first define a Fourier transformation
\begin{align}
\hat{G}(\boldsymbol{x},\omega;\boldsymbol{x}_0,t_0) = \int_{-\infty}^\infty G(\boldsymbol{x},t;\boldsymbol{x}_0,t_0) \exp(-i \omega t)\,\mathrm{d}t \label{eq:fouriertransform},\\
G(\boldsymbol{x},t;\boldsymbol{x}_0,t_0) = \frac{1}{2\pi} \int_{-\infty}^\infty \hat{G}(\boldsymbol{x},\omega;\boldsymbol{x}_0,t_0) \exp(i \omega t)\,\mathrm{d}\omega\label{eq:inversefouriertransform}.
\end{align}
Using (\ref{eq:fouriertransform}), (\ref{eq:waveequviscousperturbedgreensfunc}) can now be transformed into a Helmholtz equation
\begin{align}
-\omega^2 \hat{G}(\boldsymbol{x},\omega;\boldsymbol{x}_0,t_0) - \left(\text{St}^2+\frac{i \omega}{\text{Re}_a}\right)\nabla^2 \hat{G}(\boldsymbol{x},t;\boldsymbol{x}_0,t_0) = \delta(\boldsymbol{x}-\boldsymbol{x}_0)\exp(-i\omega t_0)\label{eq:greensfunctionwaveequationviscidfouriertransform},
\end{align}
where $i$ is the imaginary unit. A general solution to this equation can be found by expanding the Green's function into eigenfunctions, resulting in 
\begin{align}
G(r,\theta,\phi,t;r_0,\theta_0,\phi_0,t_0) = \sum_{\text{nlm}} \psi_{\text{nl}}^\text{m}(r,\theta,\phi)\psi_{\text{nl}}^\text{m}(r_0,\theta_0,\phi_0)\frac{1}{2\pi}\int_{-\infty}^{\infty} \left(\frac{\exp(i \omega(t-t_0))}{\text{St}^2\beta_{\text{nl}}^2 + \frac{i \omega \beta_{\text{nl}}^2}{\text{Re}_a} - \omega^2}\right)\,\mathrm{d}\omega,
\label{eq:greensfunctionviscidtransformed}
\end{align}
where $\psi_{\text{nl}}^\text{m}(r,\theta,\phi)$ are the eigenfunctions of the spherical Helmholtz equation
\begin{align}
\psi_{\text{nl}}^\text{m}(r,\theta,\phi) = \frac{\sqrt{2}j_{\text{l}}\left(\beta_{\text{nl}}r\right)Y_{\text{l}}^{\text{m}}(\theta,\phi)}{j_{\text{l}+1}(\beta_{\text{nl}})},
\end{align}
$j_{\text{l}}$ are the spherical Bessel functions, $Y_{\text{l}}^{\text{m}}$ are the spherical harmonics and $\beta_{\text{nl}}$ are the zeros of the spherical Bessel functions. To evaluate the inverse Fourier transform (\ref{eq:inversefouriertransform}), we use complex contour integration (see figure \ref{fig:complexintegralviscid}). It can be shown that the contribution of the arc is zero in the limit where the contour radius $a \to \infty$. Furthermore, there should be no response of an impulse released at $t_0$ at earlier times $t<t_0$ (causality condition). To this end, we pick the Jordan curve illustrated in figure \ref{fig:complexintegralviscid}a for $t>t_0$ and the curve of figure \ref{fig:complexintegralviscid}b for $t<t_0$, in the limit $a \to \infty$. A closed form expression of the Green's function is now given by
\begin{equation}
G(\boldsymbol{x},t;\boldsymbol{x}_0,t_0) = \sum_{\text{nlm}} \psi_{\text{nl}}^\text{m}(r,\theta,\phi)\psi_{\text{nl}}^\text{m}(r_0,\theta_0,\phi_0)\exp(-\kappa_{\text{nl}} (t-t_0))\frac{\sin(\eta_{\text{nl}}(t-t_0))}{\eta_{\text{nl}}}\mathcal{H}(t-t_0),
\label{eq:greensfunctionviscidfinal}
\end{equation}
where $\kappa_{\text{nl}} = \frac{\beta_{\text{nl}}^2}{2\text{Re}_a}$ and $\eta_{\text{nl}} = \frac{\sqrt{4\text{St}^2\beta_{\text{nl}}^2 - \frac{\beta_{\text{nl}}^4}{\text{Re}_a^2}}}{2}$. The resulting spatio-temporal pressure field using (\ref{eq:greensfunctiongeneralsolutionviscid}) reads (without any initial condition)
\begin{align}
&p_1(r,\theta,\phi,t) =  \sum_{\text{nl}} \beta_{\text{nl}} \frac{j_l\left(\beta_{\text{nl}} r\right)}{j_{l+1}(\beta_{\text{nl}})}\left(1-\frac{j_{l-1}(\beta_{\text{nl}})}{j_{l+1}(\beta_{\text{nl}})}\right)\frac{2l +1}{4\pi} \int_0^{2\pi}\int_0^\pi\int_0^t \exp(-\kappa_{\text{nl}} (t-t_0))\nonumber\\ &\frac{\sin(\eta_{\text{nl}}(t-t_0))}{\eta_{\text{nl}}} \mathcal{H}(t-t_0) P_l(\cos\gamma)\bigg(\text{St}^2 p_1(1,\theta_0,\phi_0,t_0) + \frac{1}{\text{Re}_a }\frac{\partial p_1(1,\theta_0,\phi_0,t_0)}{\partial t_0}\bigg)\nonumber\\ &\sin(\theta_0)\,\mathrm{d}\theta_0\,\mathrm{d}\phi_0\,\mathrm{d}t_0,
\label{eq:generalsolutionpressureviscid}
\end{align}
where $\mathcal{H}$ is the Heaviside theta function, $P_l$ are the Legendre polynomials and $\cos(\gamma) = \cos(\theta)\cos(\theta_0)+\sin(\theta)\sin(\theta_0)\cos(\phi-\phi_0)$. 
In the results section, we will use (\ref{eq:generalsolutionpressureviscid}) using a particular pressure boundary condition specified by $p_1(1,\theta_0,\phi_0,t_0)$.
\subsubsection{The velocity field}
\label{section:velocityfielddefinition}
The velocity field inside the droplet is given by (\ref{eq:momentumconservationperturbed}). Since there is no initial rotation present in the fluid and there are no rotational forces acting at later times, the velocity field remains irrotational and is given by a scalar potential. A straightforward time integral over the pressure gradient (the first term on the right hand side) based on the spherical Bessel functions (\ref{eq:generalsolutionpressureviscid}) results in a divergent series. To overcome this problem, we solve the velocity field in a different function basis. To this end, we first define the time integral over the thermodynamic pressure as the pressure impulse
\begin{align}
J_1(\boldsymbol{x},t) = \int_0^t p_1(\boldsymbol{x},t') dt'.
\label{eq:pressureimpulsedefinition}
\end{align}
The governing equation for the pressure impulse can be obtained by integration of (\ref{eq:waveequviscousperturbed}) in time
\begin{align}
\nabla^2 J_1 = \frac{1}{\text{St}^2}\frac{\partial p_1}{\partial t} - \frac{1}{\text{St}^2\text{Re}_a}\nabla^2 p_1,
\label{eq:pressureimpulsediffequation}
\end{align}
where we used that both the pressure field $p_1$ and its derivative vanish at $t=0$. It now becomes apparent that the natural basis functions for the pressure impulse are in fact harmonic functions which results in a convergent series.

The general solution for the pressure impulse inside the droplet is therefore given by
\begin{align}
J_1(\boldsymbol{x},t) =& \iiint\limits_{V}G(\boldsymbol{x};\boldsymbol{x}_0)\left[\frac{1}{\text{St}^2}\frac{\partial p_1(\boldsymbol{x}_0,t)}{\partial t} - \frac{1}{\text{St}^2\text{Re}_a}\nabla^2 p_1(\boldsymbol{x}_0,t)\right]\,\mathrm{d}V_0 \label{eq:pressureimpulsegreensfunctionintegral}\\&- \oint_S\left[G(\boldsymbol{x},\boldsymbol{x}_0)\boldsymbol{\nabla}_{\boldsymbol{x}_0} J_1(\boldsymbol{x}_0,t) - J_1(\boldsymbol{x}_0,t)\boldsymbol{\nabla}_{\boldsymbol{x}_0} G(\boldsymbol{x},\boldsymbol{x}_0)\right]\cdot\boldsymbol{n}\,\mathrm{d}S_0.\nonumber 
\end{align}
where the Green's function satisfies the Poisson equation in spherical coordinates
\begin{equation}
\nabla^2G(\boldsymbol{x};\boldsymbol{x}_0) = \delta(\boldsymbol{x}-\boldsymbol{x}_0).
\end{equation}
Completely analogous to the boundary conditions on $p_1$, the boundary condition on $J_1$ is
$J_1(1,t) = 0$, which yields
\begin{align}
G(\boldsymbol{x},\boldsymbol{x}_0) = \frac{1}{4\pi}\left(\frac{1}{\sqrt{r^2r_0^2+1-2r r_0\cos(\gamma)}}-\frac{1}{\sqrt{r^2 + r_0^2 - 2rr_0 \cos(\gamma)}}\right),
\end{align}
where $\cos(\gamma) = \cos(\theta)\cos(\theta_0)+\sin(\theta)\sin(\theta_0)\cos(\phi-\phi_0)$.
The velocity field then reads
\begin{align}
\boldsymbol{u}_1(\boldsymbol{x},t) = -\text{St}\boldsymbol{\nabla}J_1(\boldsymbol{x},t)-\left(\frac{1}{\text{Re}}+\frac{1}{\text{Re}_{\nu}}\right)\boldsymbol{\nabla}p_1,
\label{eq:momentumviscideqution}
\end{align}
where $\boldsymbol{u}_1(\boldsymbol{r},0) = \boldsymbol{0}$ and $p_1(\boldsymbol{r},0) = 0$. 

\subsection{Lattice-Boltzmann method}\label{section:lbm}
We employ an axisymmetric lattice-Boltzmann method to compare the single-phase analytic pressure field derived above to a multiphase simulation. A van-der-Waals equation of state is used, which in the vicinity of equilibrium behaves as an ideal gas. In the simulation, the ablation pressure is applied directly on the liquid-gas interface,which has a density ratio of $\sim 170$. Further details on the method can be found in \citet{Reijers2016}.

\section{Results}
\label{section:resultssection}
\subsection{Acoustic response of a droplet to the ablation pressure}
\label{section:acousticresponse}
\begin{figure}
  \centerline{\includegraphics[width=\linewidth]{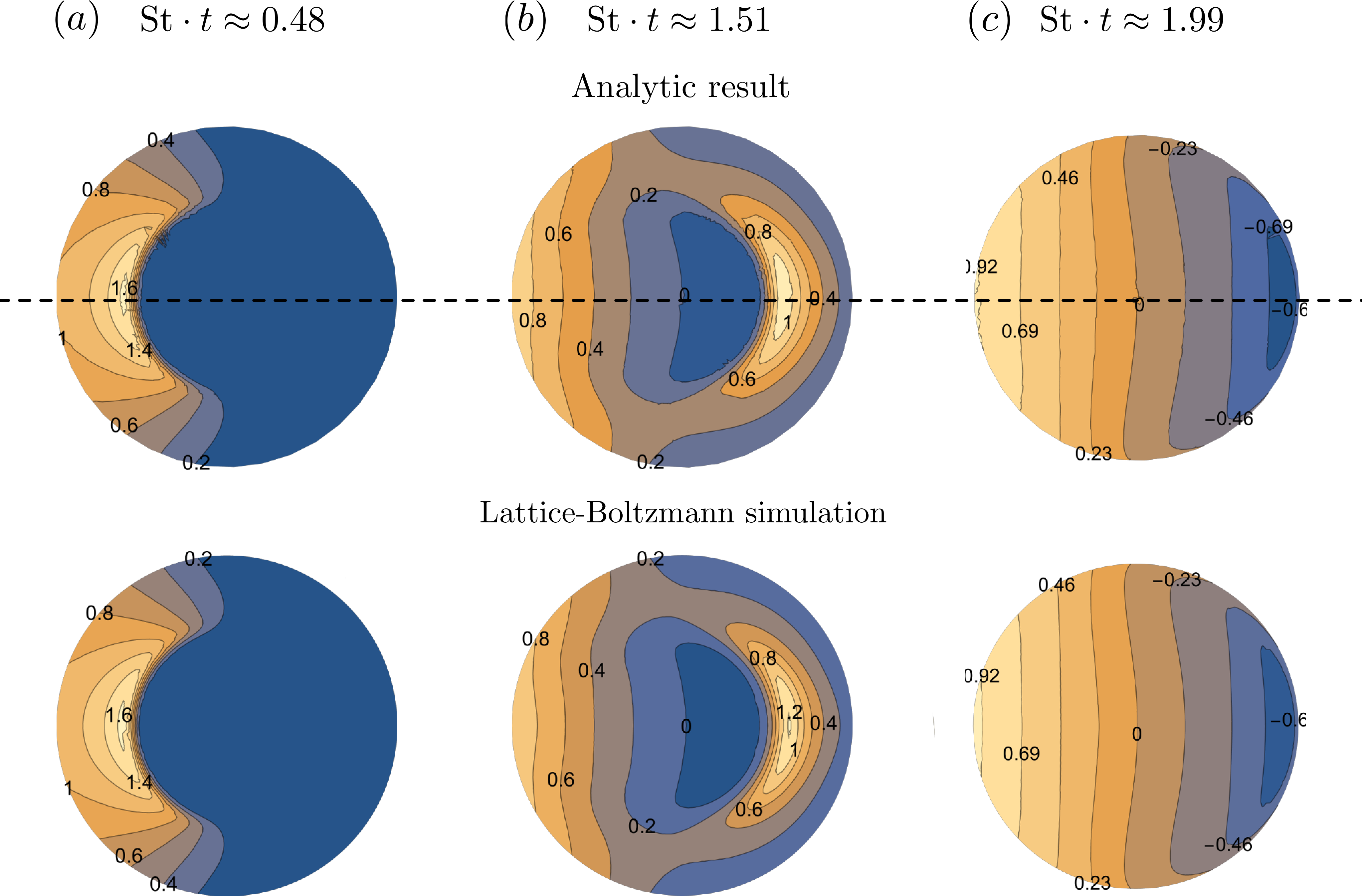}}
  \caption{The pressure field $p_1(r,\theta)$ inside the droplet at different times for an ablation pressure (\ref{eq:flatablationpressure}) impacting from the left side with a dimensionless duration $\text{St}\approx 6.02$. The top row shows the analytic results (plotted up to $r=0.95$ to restrict the amount of Fourier modes required). The bottom row shows the results of the lattice-Boltzmann simulations, which are in excellent agreement with the analytics. The black dotted line is the centerline axes used in figure \ref{fig:wavesinsidesphere}.}
\label{fig:comparisonlbmwithanalytical}
\end{figure}

\begin{figure}
  \centerline{\includegraphics[width=0.9\linewidth]{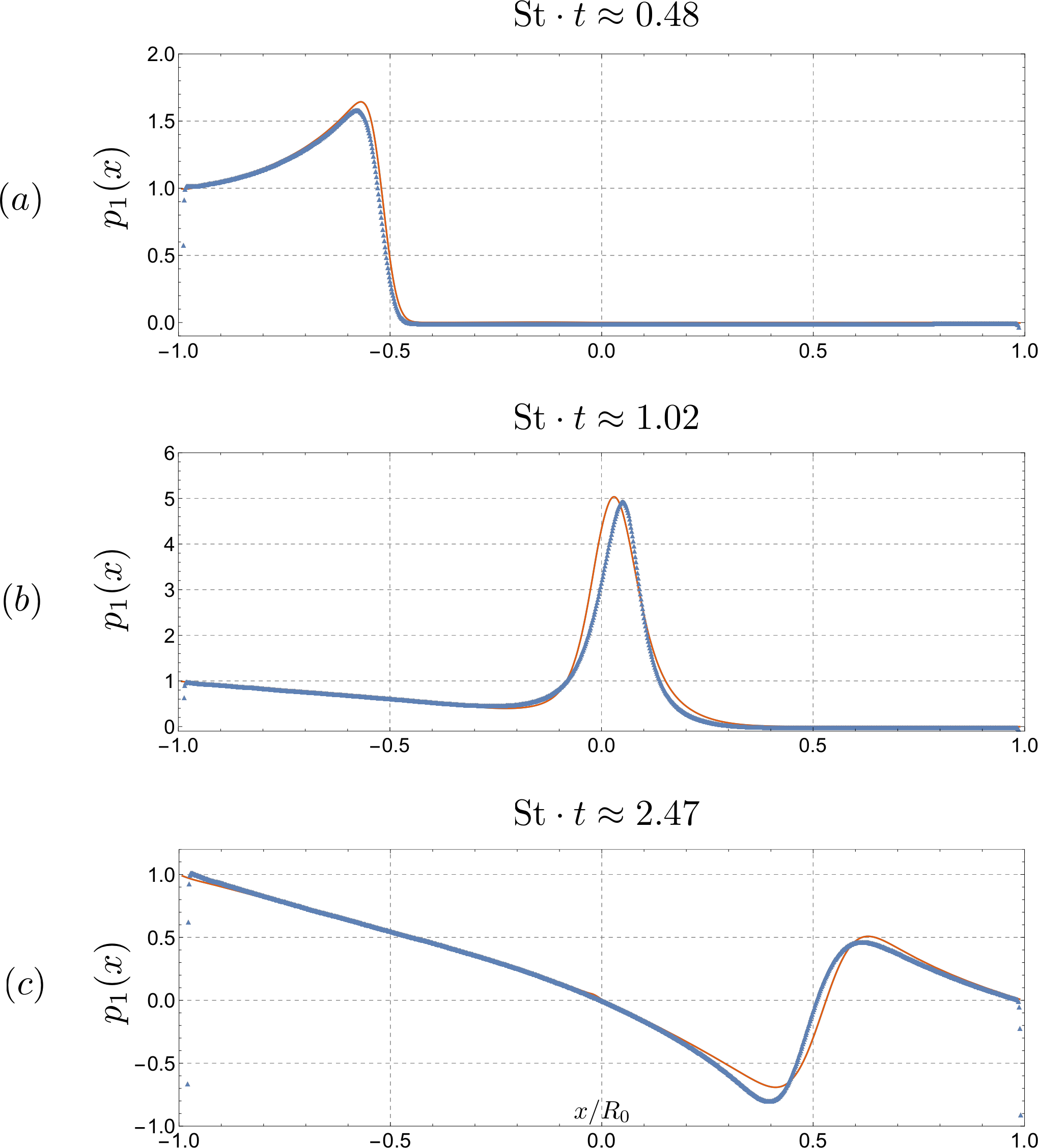}}
  \caption{Comparison between the analytic pressure $p_1(r,0)$ (red curve) and the lattice-Boltzmann simulation (blue curve) along the centerline (see figure \ref{fig:comparisonlbmwithanalytical}) at different times for $\text{St} = 6.02$. a) For $\text{St}\cdot t=0.48$ the pulse is still going on, b) for $\text{St}\cdot t=1.02$ the pulse has just finished, and c) for $\text{St}\cdot t=2.47$ the waves have reflected on the right interface of the drop and is traveling back towards the left.}
 \label{fig:wavesinsidesphere}
\end{figure}

We consider the impact of a uniform laser-beam profile on the left side of the droplet \citep{Gelderblom2016}
\begin{equation}
p_1(1,\theta,\phi,t) = \cos(\theta)\mathcal{H}\left(\frac{\pi}{2}-\theta\right)\mathcal{H}(1-t),
\label{eq:flatablationpressure}
\end{equation}
where $\mathcal{H}$ is the Heaviside function that restricts the pressure profile to the illuminated side of the droplet and limits the duration of the ablation pressure. This pressure pulse profile (\ref{eq:flatablationpressure}) will be used in all results presented below. The analytic pressure field (\ref{eq:generalsolutionpressureviscid}) subject to (\ref{eq:flatablationpressure}) is given by 
\begin{align}
&p_1(r,\theta,\phi,t) = \text{St}^2 \sum_{\text{nl}} \beta_{\text{nl}} \frac{j_l\left(\beta_{\text{nl}} r\right)}{j_{l+1}(\beta_{\text{nl}})}\left(1-\frac{j_{l-1}(\beta_{\text{nl}})}{j_{l+1}(\beta_{\text{nl}})}\right)\frac{2l +1}{4\pi} \int_0^{2\pi}\int_0^\pi\int_0^t \exp(-\kappa_{\text{nl}} (t-t_0))\nonumber\\ &\frac{\sin(\eta_{\text{nl}}(t-t_0))}{\eta_{\text{nl}}} \mathcal{H}(t-t_0)\mathcal{H}(1-t_0) P_l(\cos\gamma)\cos(\theta_0)\mathcal{H}(\frac{\pi}{2}-\theta_0) \sin(\theta_0)\,\mathrm{d}\theta_0\,\mathrm{d}\phi_0\,\mathrm{d}t_0
\label{eq:generalsolutionpressureviscidflatkick}
\end{align}
for $t<1$. 

In figure \ref{fig:comparisonlbmwithanalytical} we show a comparison between (\ref{eq:generalsolutionpressureviscidflatkick}) and the lattice-Boltmzann simulations at different times, for $\text{St} = 6.02$ and $\text{Ma} \ll 1$. In order to obtain the analytic plots we used $\text{Re}_a\sim 200$, which will be used for all results in the remainder of this paper. Initially, the pressure disturbance on the surface of the droplet sends out a radially expanding wave for all source points on the boundary inside the droplet (figure \ref{fig:comparisonlbmwithanalytical}a) which then propagates (figure \ref{fig:comparisonlbmwithanalytical}b) to the right side  (figure \ref{fig:comparisonlbmwithanalytical}c). During the propagation, the superposition of all the waves inside the droplet gives rise to a non-trivial pressure distribution, see figure \ref{fig:comparisonlbmwithanalytical}b and \ref{fig:comparisonlbmwithanalytical}c. Note that a negative value for $p_1$ does not necessarily mean a negative pressure since the total pressure is given by (\ref{eq:hydrodynamicquantities}). The figures show a good qualitative agreement between the analytic model (top row) and the simulated droplet (bottom row). 

A quantitative comparison of the pressure profiles along the centerline is given in figure \ref{fig:wavesinsidesphere}. Here, we plotted the pressure field as it passes through the center of the droplet (figure \ref{fig:wavesinsidesphere}b) and after the reflection on the right interface (figure \ref{fig:wavesinsidesphere}c). We observe good quantitative agreement between the analytic results and the simulation, also after wave reflection (figure \ref{fig:wavesinsidesphere}c) which confirms the validity of boundary condition (\ref{eq:boundaryconditionfreesurface}). In the non-ideal equation of state of the lattice-Boltzmann method, the sound speed is not constant but depends on the local pressure. This could lead to small discrepancies in comparison to the analytic model. Furthermore, in the simulation a small amount of acoustic energy could be transmitted to the gas phase when a pressure wave hits the interface.

\subsection{The effect of the pulse duration on the droplet response}
\begin{figure}
  \centerline{\includegraphics[width=0.9\linewidth]{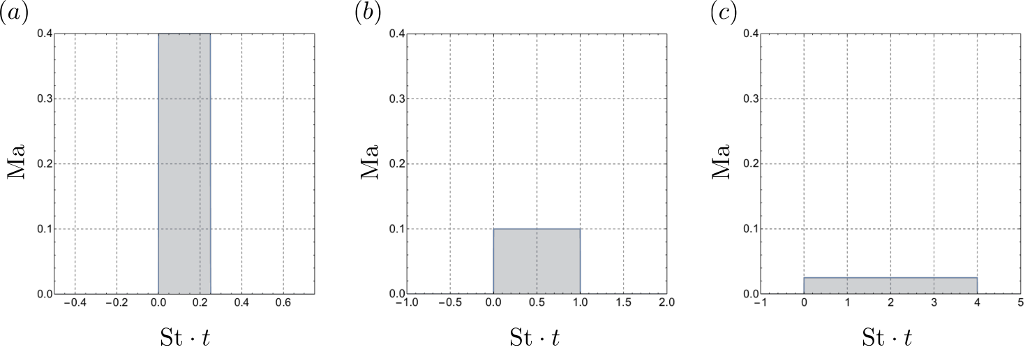}}
  \caption{Three ablation-pressure pulses of different duration but constant impulse ($\text{Ma}\text{St} = 0.1$). (a) a short pulse with a duration $\text{St} = 0.25$ and amplitude $\text{Ma} = 0.4$, (b) an intermediate pulse duration $\text{St} = 1$ and $\text{Ma}=0.1$ and (c) a long pulse duration $\text{St}=4$ and $\text{Ma}=0.025$.}
 \label{fig:differenttypesofablation}
\end{figure}
We now address how the droplet response depends on the ablation pressure amplitude ($\text{Ma}$) and duration ($\text{St}$), while the total momentum transfer to the droplet remains constant. To this end, we compare the droplet response to the three types of pulses that are illustrated in figure \ref{fig:differenttypesofablation}. In the first case (figure \ref{fig:differenttypesofablation}a) the duration of the ablation pressure is much smaller than the time it takes for a pressure wave to travel through the droplet ($\text{St} = 0.25$, $\text{Ma} = 0.4$). In the second case (figure \ref{fig:differenttypesofablation}b) the duration of the pulse is exactly equal to the time it takes to travel over a distance of one droplet radius ($\text{St}=1$, $\text{Ma}=0.1$). Finally (figure \ref{fig:differenttypesofablation}c) defines a pulse duration that is much longer than the acoustic time scale of the droplet ($\text{St}=4$, $\text{Ma}=0.025$). In all three cases, the total momentum transfer to the droplet is constant and equal to $\text{St}\text{Ma} = 0.1$.
Below, we discuss the differences in the pressure field (\S\ref{subsection:pressurefield}), pressure impulse field and velocity field inside the droplet (\S\ref{subsection:velocityfield}) and eventually the droplet deformation dynamics (\S\ref{subsection:deformation}) for these three different pulses.

\subsubsection{Pressure field}\label{subsection:pressurefield}
\begin{figure}
  \centerline{\includegraphics[width=0.9\linewidth]{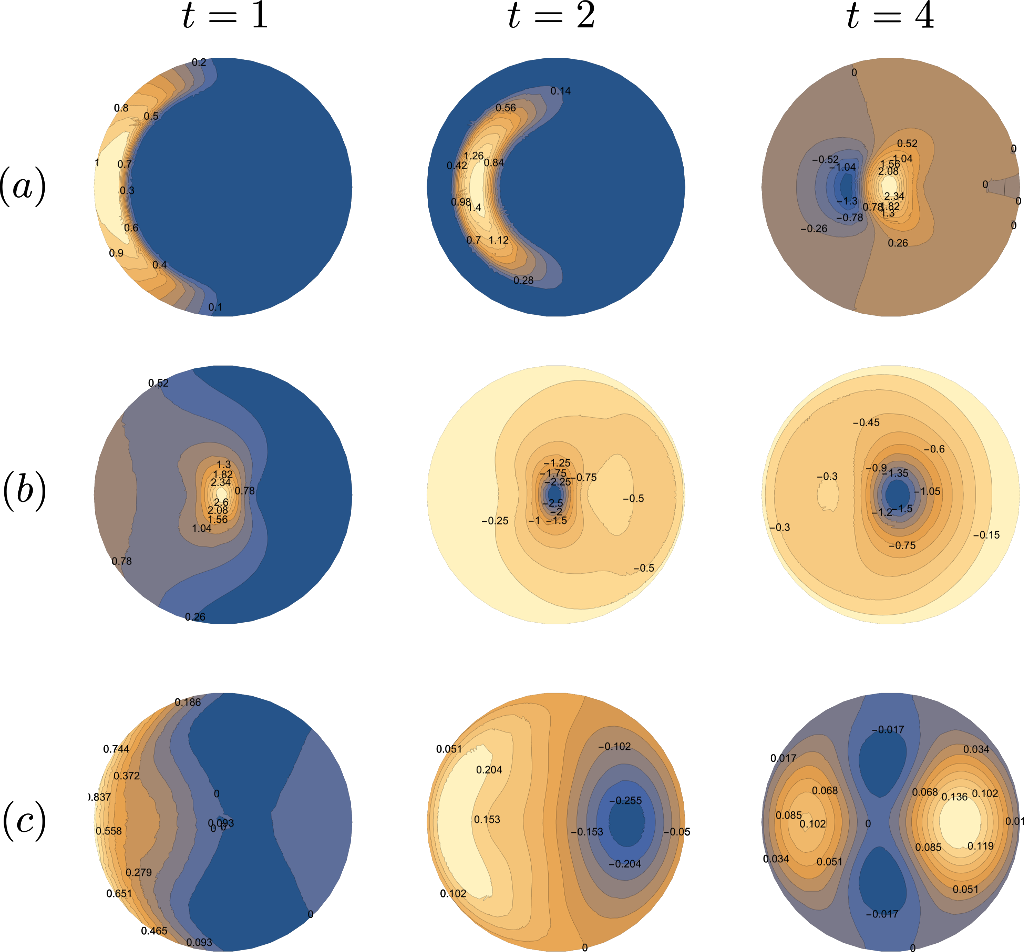}}
  \caption{The pressure field $p_1$ inside the droplet for the three different illustrated in figure \ref{fig:differenttypesofablation}: (a) $\text{St}=0.25$, (b) $\text{St} = 1$ and (c) $\text{St}=4$. The results are depicted for different times: at the end of the pulse ($t=1$), at two times the pulse duration ($t=2$) and at four times the pulse duration ($t=4$). Note that the color bar scale is not fixed and that all fields are scaled with $p_e$.}
\label{fig:wavesinspheredifferentcases}
\end{figure}

Figure \ref{fig:wavesinspheredifferentcases} shows the spatio-temporal pressure field inside the droplet that is induced by the three pressure pulses discussed in figure \ref{fig:differenttypesofablation}. For a short pulse duration, most of the pressure field is initially (i.e.~ at $t = 1$) localized in a small compression zone (figure \ref{fig:wavesinspheredifferentcases}a). We note that when $t=1$, all plots are drawn exactly after the pulse in figure  \ref{fig:wavesinspheredifferentcases}. This zone is the result of the superposition of radial compression waves emitted from source points on the interface. During the propagation ($t=2$), the superposition of these waves leads to a highly compressed spot in the center, which is clearly visible at $t=4$. At later times (not shown in the figure) the compression waves reach the right interface of the droplet where they reflect and give rise to an expansion zone. Meanwhile, wave reflections continuously occur on the left interface during the pulse. The superposition of these reflected waves leads to a expansion zone close to the left interface that is clearly visible at $t=4$, see the negative pressure zone. We note that the absolute pressure is not negative, since the absolute pressure is given by (\ref{eq:hydrodynamicquantities}).

The pressure field for intermediate pulse duration is illustrated in figure \ref{fig:wavesinspheredifferentcases}b. At the end of the pulse ($t=1$) the waves have travelled a distance $R$. Again a compression zone is created in the center, followed by an expansion zone ($t=2$). At $t=4$, all waves have at least reflected once on the interface of the droplet which gives rise to another large expansion zone. For a long pulse (figure \ref{fig:wavesinspheredifferentcases}c) the pressure field has spread over the entire droplet. The superposition of all compression and expansion waves lead to a non-trivial field that consists of compression and expansion zones.

To summarize, we observe more localized fluctuations in the pressure field directly after a short pulse as compared to longer pulses. As we will demonstrate below, these fluctuations have an important effect on how the impulse is distributed over time and hence on the resulting velocity field inside the droplet.

\subsubsection{Pressure impulse and velocity fields}\label{subsection:velocityfield}
\begin{figure}
  \centerline{\includegraphics[width=0.9\linewidth]{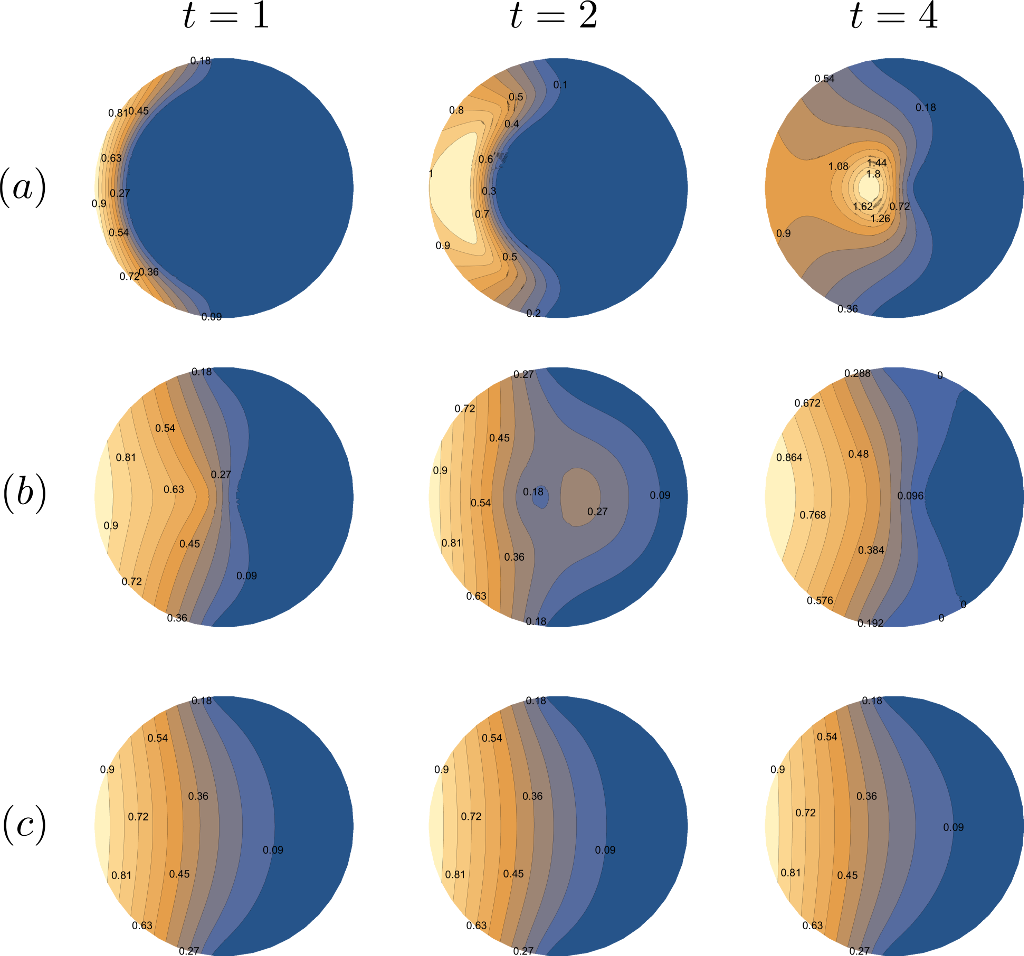}}
  \caption{The pressure impulse field $J_1$ inside the droplet for the three different pulse durations illustrated in figure \ref{fig:differenttypesofablation}: (a) $\text{St}=0.25$, (b) $\text{St} = 1$ and (c) $\text{St}=4$. The results are depicted for different times: at the end of the pulse ($t=1$), at two times the pulse duration ($t=2$) and at four times the pulse duration ($t=4$). Note that the color bar scale is not fixed and that all fields are scaled with $p_e\tau_e$.}
\label{fig:pressureimpulseplots}
\end{figure}
Figure \ref{fig:pressureimpulseplots} shows the spatio-temporal pressure impulse field (\ref{eq:pressureimpulsedefinition}) inside the droplet for the three pulse durations. To obtain the plots, we evaluate (\ref{eq:pressureimpulsegreensfunctionintegral}) numerically. This scalar field is an important field in our analysis, since it describes the spatio-temporal distribution of momentum inside the droplet and the velocity field (\ref{eq:momentumviscideqution}) is derived from it, as discussed in \S \ref{section:velocityfielddefinition}. Note that in all cases the total momentum inside the droplet is constant as soon as the pulse is over, at $t=1$, while the distribution of the momentum can still change in time.  

For a short pulse duration the momentum distribution changes significantly in time, see figure \ref{fig:pressureimpulseplots}a. Initially all momentum is concentrated on the left side of the droplet, while it redistributes itself throughout the droplet at later times. As we will show below, this localized momentum distribution results in a stronger interface deformation for short pulses. As the pulse duration increases (figure \ref{fig:pressureimpulseplots}b) the time variation of the momentum distribution becomes smaller. This effect is most prominent for long pulses (figure \ref{fig:pressureimpulseplots}c) where the momentum distribution is almost constant after $t=1$. In the limit $\text{St}\to\infty$ (and consequently $\text{Ma}\to 0$ to keep the impulse finite) the pressure impulse is constant in time which corresponds to an incompressible flow.
\begin{figure}
  \centerline{\includegraphics[width=0.9\linewidth]{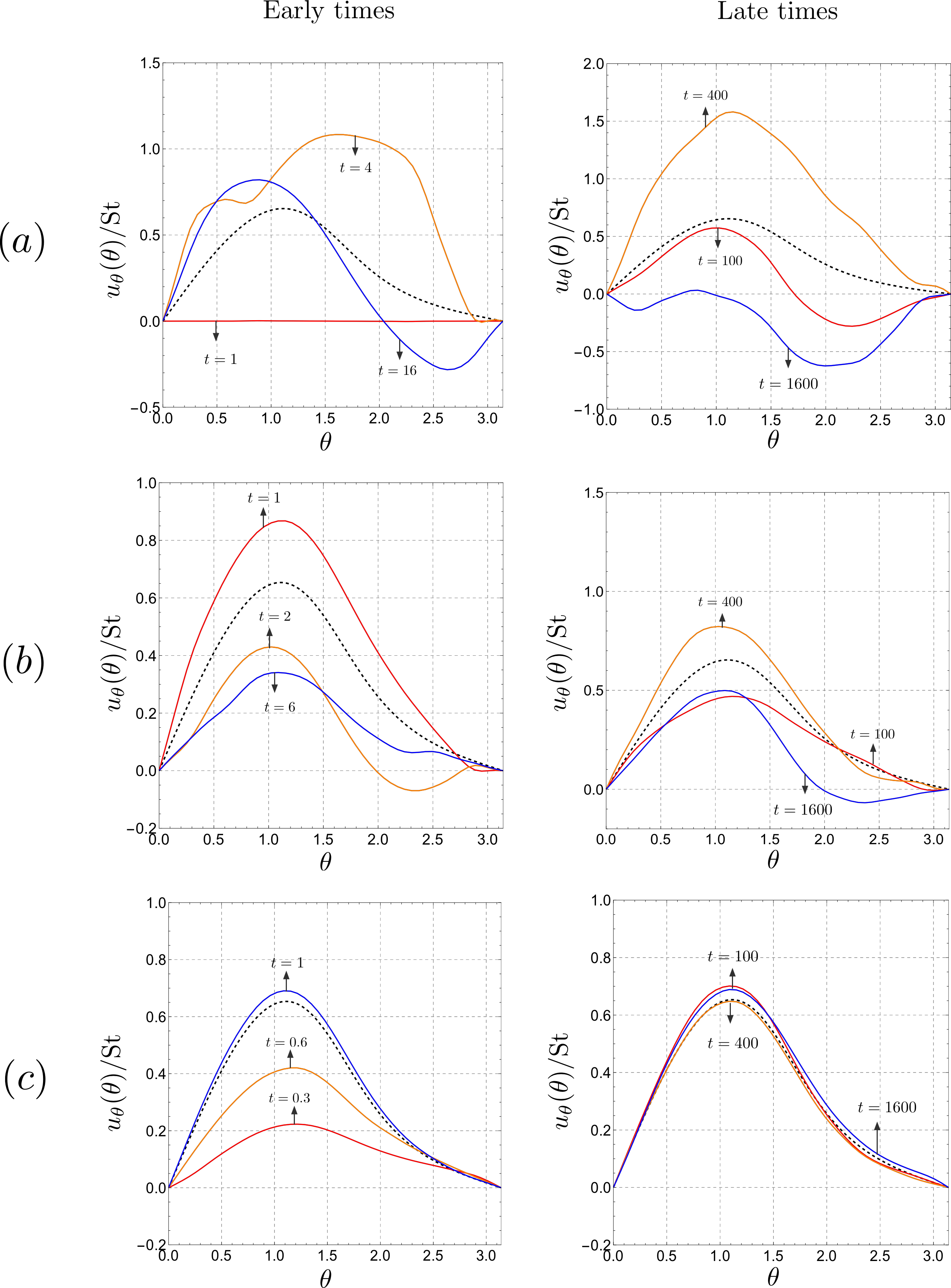}}
  \caption{The $\theta$-component of the velocity field inside the droplet at $r=0.5$. The left column represents early times ($t \sim 1$) while the right column represents late times ($t \gg 1$). (a) A short pulse duration $\text{St}=0.25$, (b) an intermediate pulse duration $\text{St} = 1$ and (c) a long pulse duration $\text{St}=4$. The black dashed line denotes the incompressible velocity field by \citet{Gelderblom2016}.}
\label{fig:velocityfielddifferentcases}
\end{figure}

The velocity field (\ref{eq:momentumviscideqution}) derived from the pressure impulse is plotted in figure \ref{fig:velocityfielddifferentcases}, where we show the $\theta-$ component for $r=0.5$ at different times (solid lines). For comparison, the incompressible velocity field as derived in  \citet{Gelderblom2016} is plotted as the black dashed line. When the pulse duration is short (figure \ref{fig:velocityfielddifferentcases}a) the velocity field at $r=0.5$ for $t=1$ is zero, since the momentum has not yet propagated far enough into the droplet. As time progresses, velocity fluctuations become apparent and even for times long after the pulse ($t\gg 1$, right panel) they are nowhere near the incompressible solution. Figure \ref{fig:velocityfielddifferentcases}b shows the velocity field for an intermediate pulse duration. Here, the velocity field fluctuates around the incompressible solution. However, the amplitude of these fluctuations are large. For the longest pulse the velocity gradually builds up (figure \ref{fig:velocityfielddifferentcases}c, left panel). At later times (right panel) the velocity field shows only tiny fluctuations around the incompressible solution.  

\subsubsection{Droplet deformation}\label{subsection:deformation}
Finally, we turn to the question how the droplet deformation is affected by the duration of the ablation pressure pulse.
To make a prediction for the droplet deformation, we use the velocity field at the droplet surface. Strictly speaking, the analytic solution (\ref{eq:momentumviscideqution}) is derived for a constant spherical domain. However, we can obtain a first order approximation of the droplet shape at early times, i.e.~ when the deviations from a spherical shape are still small, by advecting material points on the interface as described in \citet{Gelderblom2016}.

\begin{figure}
  \centerline{\includegraphics[width=0.95\linewidth]{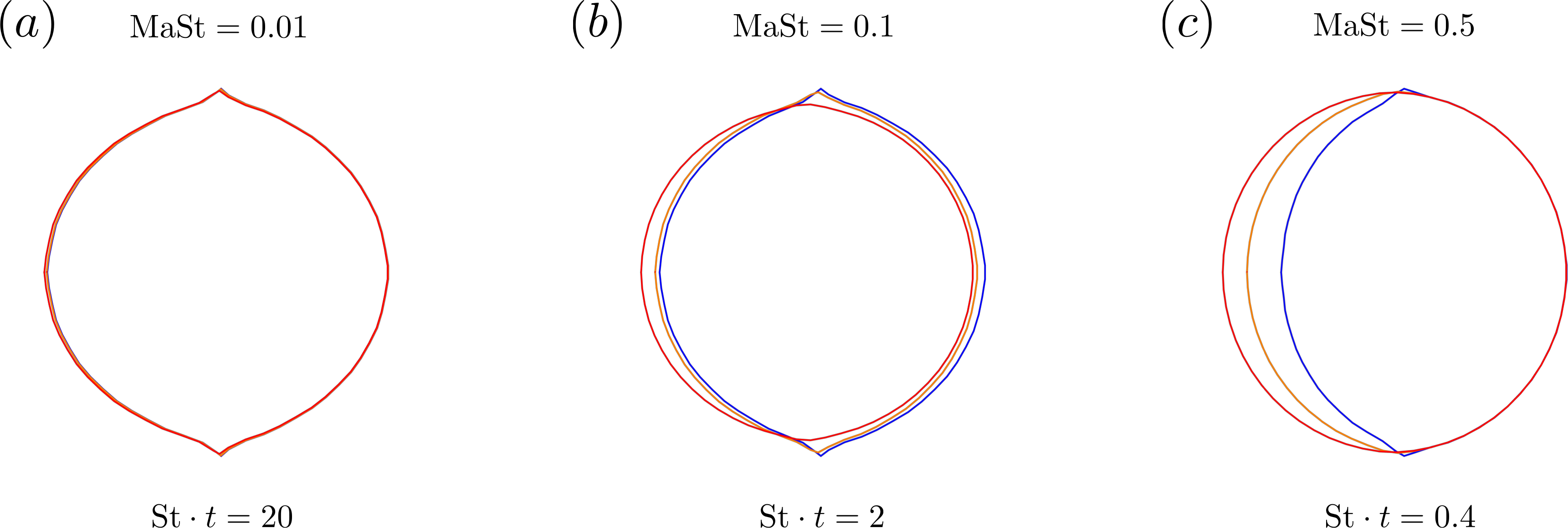}}
  \caption{Contour plots of the droplet deformation for the three different pulse durations: $\text{St}=0.25$ (blue), $\text{St}=1$ (orange) and $\text{St}=4$ (red). (a) For $\text{Ma}\text{St} = 0.01$ all cases give rise to identical deformation behavior, for (b)  $\text{Ma}\text{St} = 0.1$ we see discrepancies arising between the three pulse durations which aggravate in (c)  $\text{Ma}\text{St} = 0.5$ where we observe a clear influence of the pulse duration on the droplet deformation. Note that for each impulse the contours are sketched at a different absolute time to be able to clearly illustrate the deformations. }
\label{fig:dropdeformationsphericalrep}
\end{figure}
\begin{figure}
  \centerline{\includegraphics[width=0.95\linewidth]{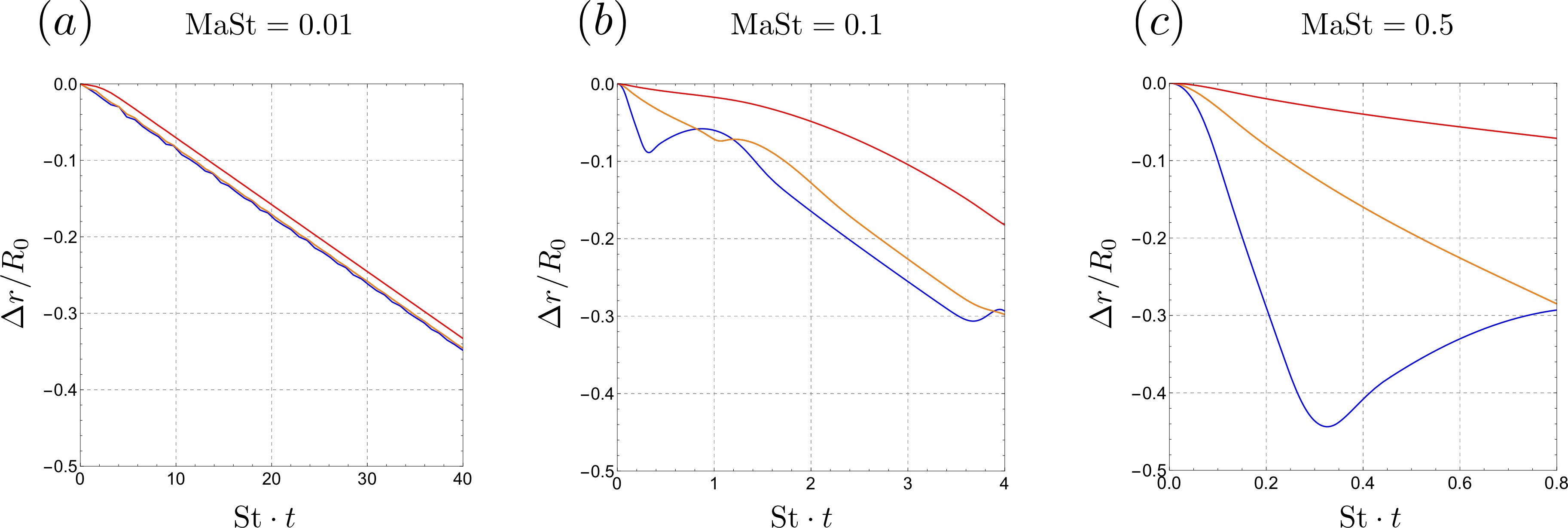}}
  \caption{The displacement of the droplet interface $\Delta r/R$ at the axis of impact ($r=1$, $\theta = 0$) for the three different pulse durations: $\text{St}=0.25$ (blue), $\text{St}=1$ (orange) and $\text{St}=4$ (red). (a) $\text{Ma}\text{St} = 0.01$, (b) $\text{Ma}\text{St} = 0.1$ (c)  $\text{Ma}\text{St} = 0.5$.}
\label{fig:dropdeformationanalytic}
\end{figure}

We compare the effect of the three different pulse durations illustrated in figure \ref{fig:differenttypesofablation} on the droplet deformation. The droplet deformation is not only determined by the pulse duration, but also by the pulse amplitude. We therefore additionally consider three different momentum transfers to the droplet: $\text{St}\text{Ma} = 0.01$,  $\text{St}\text{Ma} = 0.1$ and  $\text{St}\text{Ma} = 0.5$. The acoustical Mach number (or the product $\text{St}\text{Ma}$) now becomes an additional parameter, because we want to quantify the actual differences in deformation. So far, we always scaled out this amplitude dependency (\ref{eq:scaling}). However, a difference in amplitude now gives a stronger or weaker deformation. In figure \ref{fig:dropdeformationsphericalrep} we show contours of the droplet deformation for the three different momentum transfers and compare the influence of the pulse duration. In figure \ref{fig:dropdeformationanalytic} we quantify the interface displacement of the droplet at the axis of impact $\Delta r/R$ in time. 

When the momentum transfer to the droplet is small ($\text{Ma}\text{St} = 0.01$, figures \ref{fig:dropdeformationsphericalrep}a and \ref{fig:dropdeformationanalytic}a), we need to go to late times ($\text{St}\cdot t \gg 1$) in order to observe a significant deformation. On the acoustic time scale $R/c$ however, the droplet deformation only shows tiny fluctuations around a static shape. In other words, the fluctuations in the velocity field due to the pressure waves are negligible and the velocity field is to a good approximation incompressible. As a result, on late times the droplet deformation for the three different pulse durations is indistinguishable.

When the amount of momentum transfer to the droplet is increased (figures \ref{fig:dropdeformationsphericalrep}b and \ref{fig:dropdeformationanalytic}b), interface deformations become apparent at earlier times. For the shortest two pulses all momentum was transferred into the droplet at the time of the plot, while for the long pulse the ablation pressure is still acting on the droplet surface. Therefore, the contour of the longest pulse (red curve) is lagging behind compared to the contour of the shorter pulses (yellow and blue curves). Furthermore, for shorter pulses the droplet interface compression is followed by an expansion directly after the pulse which is visible in figure \ref{fig:dropdeformationanalytic}b. Hence, the droplet deformation for short pulses is now clearly a non-monotonous function of time. As $\text{Ma}\text{St}$ is further increased, the droplet deformation shows even larger fluctuations around a shape that is also globally deforming, see figures \ref{fig:dropdeformationsphericalrep}c and \ref{fig:dropdeformationanalytic}c.

These strong deformations invalidate the assumption that the droplet remain stationary during the pulse. Although figures \ref{fig:dropdeformationsphericalrep}c and \ref{fig:dropdeformationanalytic}c give a first order estimate of the deformations that are to be expected, for a quantitative prediction one has to solve the fields in the deformed geometry. In this regime we therefore anticipate a strong influence of the pulse duration on the eventual droplet-shape evolution at later times.

\section{Discussion \& conclusion}
The droplet deformation resulting from a laser-induced ablation pressure pulse is studied analytically in the regime where the pulse duration is of the order of the acoustic time scale and the pressure fluctuations are small. The resulting momentum change of the droplet is determined by the pressure pulse amplitude and duration or, in dimensionless form, the acoustic Mach number $\text{Ma}$ and the acoustic Strouhal number $\text{St}$. We examined the effect of changing $\text{St}$ (i.e. shortening the pulse duration) on the droplet response while keeping the total impulse transferred to the droplet constant.

The pressure, pressure impulse and velocity fields inside the droplet are studied as function of $\text{St}$ at constant impulse $\text{St}\text{Ma}$. To keep the analysis simple we used a cosine-shaped ablation pressure profile on the surface of the droplet together with a step function to limit the ablation pressure in time and space. To get a first order estimate of the droplet deformation in time we advected material points on the surface.

In the regime where $\text{St}\text{Ma} \ll 1$, the droplet deformation is independent of $\text{St}$ and no significant changes in the deformation were observed for shorter pulses. When $\text{St}$ is large, the flow inside the droplet may be considered incompressible since the pressure impulse field is approximately constant in time. By contrast, when $\text{St}\ll 1$ the flow inside the droplet is compressible. However, on the deformation time scale the compressible effects average out and the droplet behaves as if it were incompressible. Therefore, the incompressible model by \citet{Gelderblom2016} can be used to describe the deformation dynamics in this regime.

Significant differences in deformation arise when $\text{St}\text{Ma}\lesssim 1$. When $\text{St}\gg 1$ the flow is incompressible however now the droplet deforms significantly during the pulse. When $\text{St}\ll 1$ all momentum is localized in a small shell close to the illuminated side of the droplet directly after the pulse. This results in a high acceleration of the interface and consequently a compression of the fluid that leads to a different deformation compared to the case where the pulse duration is long ($\text{St}\gg 1$). The droplet deformation in this regime is therefore strongly dependent on the pulse duration. In practice, to study droplet deformation resulting from femto-, pico-, nano-second laser pulses in the plasma-mediated ablation regime (i.e. short ablation pressure pulses $\text{St}\lesssim 1$) at high energy (such that $\text{St}\text{Ma}\lesssim 1$), droplet compressibility needs to be taken into account.

In the regime where $\text{Ma}\sim 1$, the linear approximation of the proposed analytic model breaks down. The flow is governed by shock-waves, cavitation phenomena, nonlinear viscous damping and rapid interface acceleration, which result in a highly nonlinear droplet response. We argue however that the compressible model can be used as a starting point to identify likely cavitation spots and study first order droplet deformation, since shock fronts first need to develop in time. A more detailed understanding the droplet deformation in this regime requires numerical simulations and is topic of future work.
\section{Acknowledgements}
We are grateful to Alexander Klein, Detlef Lohse, Andrea Prosperetti, Federico Toschi and Michel Versluis for valuable discussions. This work is part of an Industrial Partnership Programme of the Foundation for Fundamental Research on Matter (FOM), which is financially supported by the Netherlands Organization for Scientific Research (NWO). This research programme is co-financed by ASML.

\bibliographystyle{jfm}
\bibliography{references}

\end{document}